
%
%

\tolerance=2000
\hbadness=2000
\overfullrule=0pt


\font\titlesf=cmssbx10 scaled \magstep2
\font\chaptsf=cmssbx10 scaled \magstep1
\font\subchaptsf=cmssbx10 scaled \magstephalf
\font\authorsf=cmss10 scaled \magstep1
\font\sf=cmss10
\font\pl=cmssq8 scaled \magstep1

\font\sc=cmcsc10

\def\a{\alpha}
\def\b{\beta}

\def\coth{{\rm coth\,}}
\def\d{\delta}

\def\det{{\rm det\,}}
\def\eps{\varepsilon}

\def\frac#1#2{{\textstyle {#1\over #2}}}
\def\g{\gamma}
\def\h#1{{\cal #1}}
\def\l{\lambda}
\def\m{\mu}

\def\n{\nu}
\def\na{\nabla}
\def\om{\omega}

\def\RR{\rm I\!R}

\def\s{\sigma}
\def\sinh{{\rm sinh\,}}

\def\log{{\rm log\,}}

\def\sq{\mathchoice{\square{6pt}}{\square{5pt}}{\square{4pt}}
    {\square{3pt}}}
\def\square#1{\mathop{\mkern0.5\thinmuskip\vbox{\hrule\hbox{\vrule
    \hskip#1 \vrule height#1 width 0pt \vrule}\hrule}\mkern0.5\thinmuskip}}

\def\hb{\hfill\break}

\def\ltextindent#1{\hbox to \hangindent{#1\hss}\ignorespaces}

\def\today{\ifcase\month\or January\or February\or March\or April\or May
    \or June\or July\or August\or September\or October \or November
    \or December \fi\space\number\day, \number\year}
\def\heute{\number\day. {\ifcase\month\or Januar\or Februar\or M\"arz
    \or April\or Mai\or Juni\or Juli\or August\or September\or Oktober
    \or November\or Dezember\fi} \number\year}

\def\rightheadline{\it\title\qquad\hfill\rm\folio}
\def\leftheadline{\rm\folio\hfill\it\qquad\author}
\headline={\ifnum\pageno>1{\ifodd\pageno\rightheadline\else\leftheadline\fi}
\else\fi}


\nopagenumbers
\def\author{I. G. Avramidi}
\def\title{Algebraic approach to the heat kernel}

\null
\vskip-1.5cm
\hskip6.2cm{ \hrulefill }
\vskip-.55cm
\hskip6.2cm{ \hrulefill }
\smallskip
\hskip6.2cm{{\pl \ University of Greifswald (May, 1994)}}
\smallskip
\hskip6.2cm{ \hrulefill }
\vskip-.55cm
\hskip6.2cm{ \hrulefill }
\bigskip
\hskip6.2cm{\ hep-th/9406047}
\bigskip
\hskip6.2cm{\ Revised version, August 1995}

\hskip6.2cm{\ Accepted for publication in:}

\hskip6.2cm{\sf \ Journal of Mathematical Physics}

\vfill

\centerline{\titlesf A new algebraic approach for calculating the heat kernel}
\medskip
\centerline{\titlesf in quantum gravity}
\bigskip

\centerline{{\authorsf I. G. Avramidi}
\footnote{$^{a)}$}{Alexander von Humboldt Fellow}
\footnote{$^{b)}$}{On leave of absence from the Research Institute for Physics,
Rostov State University, Stachki 194, Rostov-on-Don 344104, Russia}
\footnote{$^{c)}$}{E-mail: avramidi@math-inf.uni-greifswald.d400.de}}
\smallskip
\noindent
\centerline{\it Department of Mathematics, University of Greifswald}
\centerline{\it Jahnstr. 15a, 17489 Greifswald, Germany}

\vfill
{\narrower
\noindent
It is shown that the heat kernel operator for the Laplace operator on any
covariantly constant curved background, i.e. in symmetric spaces, may be
presented in form of an averaging over the Lie group of isometries with some
nontrivial measure. Using this representation the heat kernel diagonal, i.e.
the heat kernel in coinciding points is obtained. Related topics concerning the
structure of symmetric spaces and the calculation of the effective action are
discussed.
\par}
\par
\eject


\leftline{\chaptsf I. INTRODUCTION}
\bigskip
The heat kernel, a very powerful tool for investigating the effective action
in quantum field theory and quantum gravity, has been the subject of much
investigation in recent years in physical as well as in mathematical
literature
(Refs. 1-22). The subject of present investigation is the low-energy limit of
the
one-loop contribution of a set of quantized fields $\phi$
on a  $d$-dimensional
Riemannian manifold $M$ of metric $g_{\m\n}$ with Euclidean signature to the
effective action, which can best be presented using the $\zeta$-function
regularization in the form$^1$
$$
\Gamma_{(1)}=-{1\over 2}\zeta'(0) ,                       \eqno(1.1)
$$
where
 $$
\eqalignno{
&\zeta(p)=\m^{2p} {\rm Tr}\, F^{-p} =  {\m^{2p} \over \Gamma (p)}
 \int\limits_0^\infty dt\ t^{p-1} {\rm Tr}\, U(t) ,    &(1.2)\cr
&F=-\sq + Q + m^2  ,                                &(1.3)\cr
&U(t) = \exp(-t F)   ,                                   &(1.4)\cr}
$$
 with $\sq=g^{\m\n}\na_\m\na_\n$, $Tr$ meaning the functional
 trace, $\m$ being a renormparameter introduced to preserve dimensions,
 $Q(x)$ an arbitrary  matrix-valued function (potential term), $m$  a mass
parameter and
 $\na_\m$ a covariant derivative. The covariant derivative includes, in
general, not only the Levi-Civita connection but also the appropriate spin one
as well as the vector gauge connection and is determined by the commutator
$[\na_\m,\na_\n]\phi={\cal R}_{\m\n}\phi$. The Riemann curvature tensor, the
curvature of background connection and the potential term completely describe
the background metric and connection, at least locally. In the following we
will call these quantities the {\it background curvatures} or simply curvatures
and denote them symbolic by $\Re=\{R_{\m\n\a\b}, {\cal R}_{\m\n}, Q \}$.


Exact evaluation of the heat kernel $U(t)$ is obviously impossible.  Therefore,
one should make use of various approximations.
First of all, let us note the very important so called  Schwinger -  De
Witt asymptotic expansion of the heat kernel at $t\to 0$
$^{1-5}$
$$
\eqalignno{
& {\rm Tr}\, U(t) \sim (4\pi t)^{-d/2}\exp(-tm^2)
\sum\limits_{k=0}^\infty {(-t)^k\over k!} B_k   ,        &(1.5)\cr
 &B_k = \int_M dx g^{1/2} {\rm tr}b_k      .                     &(1.6)\cr}
 $$

This expansion is purely local  and does not depend, in fact,
on the global structure of the manifold.
In manifolds with boundary additional terms in $B_k$ as well as new terms
 of order $t^{-d/2+k/2}$ in form of surface integrals over the
 boundary $\partial M$ appear. For details see
Refs. 12,13, where all coefficients for arbitrary boundary conditions up to
terms of order $t^{-d/2+1}$ are calculated.
Its coefficients $b_k$ (we call them
Hadamard - Minakshisundaram - De Witt - Seeley (HMDS) coefficients)
are  local invariants built from the curvature,  the potential term
and their covariant derivatives.$^{1,5,6,14}$
 They play a very important role both in physics and mathematics
and are closely connected with various sections of mathematical
physics.$^{14,22}$
 Therefore, the calculation of HMDS-coefficients is in itself of
great importance. Various methods were used for calculating these
coefficients, beginning from the direct De Witt's method$^1$
to modern mathematical methods, which make use of pseudodifferential
operators, functorial properties of the heat kernel etc.$^{5-13}$
 Very good reviews of the calculation of the HMDS-coefficients
are given in recent papers.$^{14}$

Nowadays, in general case only the first four coefficients
are explicitly calculated. The first three coefficients were calculated in Ref.
9.
 An effective covariant technique for calculating HMDS-coefficients is
elaborated in Refs. 10, 4,
 where also the first four coefficients are computed. In the case of
scalar operators the fourth coefficient is also calculated in Ref. 11.
  Analytic approach was developed in Ref. 7,
  where a closed form for the intrinsic symbol of the resolvent parametrix was
obtained. The leading terms in all the volume coefficients $B_k$
quadratic in the background
curvatures were calculated completely independently in Refs. 15, 16.

Although the Schwinger - De Witt expansion is good  for small $t$, (viz. $ t
\Re\ll 1 $), and thereby  in  the case of massive quantized fields in weak
background fields when  $\Re \ll m^2$, it is absolutely inadequate  for
 large $t$ in strongly curved manifolds and strong background fields ($\Re \gg
m^2$).  For investigating these cases one needs some other methods.

A possibility to exceed the limits of the Schwinger - De Witt expansion is
to employ  the direct partial summation.$^2$
Namely, one can compare all the terms in HMDS-coefficients $B_k$
(1.6),  pick up the main (the largest in some approximation) terms and
sum up the corresponding partial sum. There is always a lack of uniqueness
concerned with the global structure of the manifold, when doing so.  But,
hopefully, fixing the topology, e.g. the trivial one, one can obtain a unique,
well defined, expression that would reproduce the Schwinger -De Witt
expansion, being expanded in curvature. The main advantage of such an approach
is that although the result will be {\it not exact} it will be {\it covariant}
and {\it general}.

Actually, the effective action is a covariant functional of  the metric
and depends on  the
geometry of the  manifold as a whole,  i.e.  it depends on both local
characteristics of the geometry like invariants of the curvature tensor and
its global topological structure. However, {\it we will not  investigate  in
this paper the influence  of the topology} but
concentrate our attention, as a rule,  on the {\it local effects}.
That means that we restrict ourselves to those physical problems where the
contribution of the global effects may be neglected in comparison with local
ones.
Then the
possible approximations for evaluating the effective action can be based on
the assumptions about the local behavior of the background fields, dealing
with the real physical gauge invariant variations of  the local geometry,
i.e. with the curvature invariants, but not with the behavior of the
metric and the connection which is not invariant.  Comparing the value of the
curvature with
that of its covariant derivatives one comes to two possible approximations:
i) the short-wave (or high-energy) approximation characterized by $\na\na
\Re\gg \Re \Re $ and ii) the long-wave  (or low-energy) one $\na\na \Re\ll \Re
\Re $.

The idea of partial summation was realized in short-wave approximation
for investigating the nonlocal aspects of the effective action (in
other words the high-energy limit of that) in Ref. 15,4,
 where all the terms in the HMDS-coefficients  $B_k$ with higher
derivatives (quadratic in the  curvature and potential term) are
calculated and the corresponding asymptotic expansion is summed up.
Another approach to study the high-energy limit of the effective action, so
called covariant perturbation theory, is
developed in Ref 17.

\bigskip
\bigskip
\leftline{\chaptsf II. LOW ENERGY APPROXIMATION AND ITS CONSEQUENCES}
\bigskip

The low-energy  effective action, in other words, the effective potential,
presents a very natural tool for investigating the vacuum of the theory, its
stability and the phase structure.$^{23}$
 Here only partial success is achieved and various approaches to the
problem are only outlined (see, e.g. the excellent review of Camporesi in Ref.
22 with an ample bibliography and our recent papers$^{20,21}$).

The long-wave (or low-energy) approximation is determined, as it was already
stressed above, by strong slowly varying  background fields.  This means that
the derivatives of all invariants are much smaller than the  products of the
invariants themselves. The zeroth order of this approximation corresponds to
covariantly constant background curvatures
$$
\na_\m R_{\a\b\g\d} = 0,\qquad \na_\m{\cal R}_{\a\b}=0,\qquad \na_\m Q = 0.
							\eqno(2.1)
$$

In this case the HMDS-coefficients are simply polynomials in curvature
invariants and potential term of dimension $\Re^k$ up to terms with one
or more covariant derivatives of the background curvatures $O(\na \Re)$
$$
\eqalignno{
&b_k=\sum^k_{n=0}{k\choose n}Q^{k-n}a_n + O(\na \Re) ,   &(2.2)\cr
&a_k=b_k\Big\vert_{Q = \na R = 0} = \sum R^k    .        &(2.3)\cr}
$$
Note that the commutators $[Q,{\cal R}_{\m\n}]$ are of order $O(\na\na \Re)$
and, therefore are neglected here.

Then after summing the Schwinger-De Witt expansion (1.5)
we obtain for the heat
kernel, the $\zeta$-function and the effective action
$$
\eqalignno{
&{\rm Tr}\, U(t) = \int_M dx\, g^{1/2}(4\pi t)^{-d/2}{\rm tr}\left\{
	\exp{\left(-t(m^2+Q)\right)}
	\left(\Omega(t)  + O(\na \Re)\right)\right\}, &(2.4) \cr
&\zeta(p)= \int_M dx\, g^{1/2}(4\pi)^{-d/2}{\m^{2p} \over \Gamma (p)}
	\int\limits_0^\infty dt\ t^{p-d/2-1}{\rm tr}\left\{
	 \exp{\left(-t(m^2+Q)\right)}
	\left(\Omega(t)+O(\na \Re)\right)\right\},        & \cr
	& &(2.5)\cr
&\Gamma_{(1)} = \int_M dx\, g^{1/2}\{ V(\Re)  + O(\na \Re) \} , & (2.6)   \cr}
$$
with
$$
\eqalignno{
V(\Re) = {1\over 2}(4\pi)^{-d/2}{1\over \Gamma({d\over 2}+1)}
\int\limits_0^\infty dt &\left(\log (\m^2 t)+\psi\left({d\over
2}+1\right)\right) &\cr
&\times\left({\partial \over \partial t}\right)^{{d\over 2}+1}{\rm
tr}\left\{\exp(-t(m^2+Q))\Omega(t)\right\} &(2.7)\cr}
$$
for even $d$ and
$$
V(\Re) = {1\over 2}(4\pi)^{-d/2}{1\over \Gamma({d\over 2}+1)}
\int\limits_0^\infty dt t^{-1/2}\left({\partial \over \partial
t}\right)^{{d+1\over 2}}{\rm tr}\left\{\exp(-t(m^2+Q))\Omega(t)\right\}
\eqno(2.8)
$$
for odd $d$, where
$$
\Omega(t) \sim  \sum\limits_{k=0}^\infty {(-t)^k\over k!} a_k  , \eqno(2.9)
$$
is a function of local invariants of the curvatures (but not of the potential).

It is naturally to call the functions $\Omega(t)$ and $V(\Re)$,  that do not
contain the {\it covariant} derivatives at all and so determine the zeroth
order of the heat kernel and that of the effective action, the {\it
generating function} for covariantly constant terms in HMDS-coefficients and
the {\it
effective potential} in quantum gravity respectively.

Let us note that such a definition of the effective potential is not
conventional. It differs from the definition that is often found in the
literature.$^{24}$
 What is meant usually under the notion of the effective potential
is a function of the potential term only $Q$, because it does not  contain
derivatives of the background fields (in contrast to Riemann curvature
$R_{\a\b\g\d}$ that contains second derivatives of the metric and the curvature
${\cal R}_{\m\n}$ with first derivatives of the connection).  So, e.g. in Ref.
24
 the  potential term $Q$ is summed up exactly but an expansion is
made not only in covariant derivatives but also in powers of curvatures
$R_{\m\n\a\b}$ and ${\cal R}_{\m\n}$, i.e. the curvatures are treated
perturbatively. Thereby the
validity of this approximation for the effective action  is limited to  small
curvatures ${\cal R}_{\m\n}, R_{\m\n\a\b} \ll Q $.  Such an expansion is called
`expansion of
the effective action in covariant derivatives'. Without the potential term
($Q=0$) the effective potential in such a scheme is trivial. Hence we stress
here once again, that the effective potential in our definition contains, in
fact,  {\it much more information} than the usual effective potential does
when using the `expansion in covariant derivatives'.
As a matter of fact, what we mean is the {\it low-energy limit of the
effective action} formulated in a covariant way.

Note that the conditions (2.1) are local. They determine the geometry of the
{\it locally} symmetric spaces.  However, the manifold is {\it globally}
symmetric one only in the case when it satisfies additionally some global
topological restrictions (e.g. it is sufficient if it is simply connected and
complete)
and the condition (2.1) is valid everywhere, i.e. at any point of
the manifold.$^{25,26}$

In most {\it physical } problems, the situation is radically
different. The correct setting of the problem seems to be as follows.
The low-energy  effective action depends, in general, also essentially on the
global topological properties of the space-time manifold, i.e. on the existence
of closed geodesics, boundaries or singularities that might act similarly to
boundaries.
But, as it was
noted above, we  do not  investigate  in this paper the influence  of the
 topology.  Therefore, consider a complete noncompact asymptotically flat
manifold without boundary that is homeomorphic to $\RR^d$.  Let  a  finite
not small, in general, domain of the manifold exists that is strongly curved
and quasi-homogeneous, i.e. the invariants of the curvature in this region
vary very slowly.  Then the geometry of this region is {\it locally} very
similar
to that of a symmetric space.  However one should have in mind that there are
{\it always} regions in the manifold where this condition is not fulfilled.
This is, first of all, the asymptotic Euclidean region that has small
curvature and, therefore, the opposite short-wave approximation is valid.

The general situation in correct setting of the problem is the
following.  From infinity with small curvature and possibly radiation,
where$^{17}$
 $\Re \Re \ll \na\na \Re $,  we pass on to quasi-homogeneous region where
the local properties of the manifold are close to those of symmetric spaces.
The size of this region can tend to zero. Then the curvature is nowhere large
and the short-wave approximation is valid anywhere.
If one tries to extend the limits of such region to infinity, then one has
also to analyze the topological properties. The space can be compact or
noncompact depending on the sign of the  curvature.  But first we will come
across a coordinate horizon-like singularity, although no one true physical
singularity really exists.

This construction can be  intuitively imagined as follows. Take the flat
Euclidean space $\RR^d$, cut out from it a region $M$ with some boundary
$\partial M$ and stick to it smoothly along the boundary, instead of the piece
cut
out, a piece of a curved symmetric space with the same boundary $\partial M$.
Such a construction will be homeomorphic to the initial space and at the same
time will contain a finite highly curved homogeneous region.
Let us stress, that this surgery can be always done {\it smoothly}, so that in
the region where the curved and the flat regions are joined no discontinuity in
the curvature appears that could cause the reflected waves to produce
Casimir-like effects.
By the way,
the exact effective action for a symmetric space differs from the  effective
action for built construction by a purely topological  contribution.
This fact seems to be useful when analyzing the effects of topology.

Thus the problem is to calculate  the low-energy effective action (2.7), (2.8),
i.e. the heat kernel for covariantly constant background. Although this
quantity, generally speaking, depends essentially on the topology and other
global aspects of the manifold, one can disengage oneself  from these effects
fixing the trivial topology. Since the asymptotic Schwinger - De Witt
expansion does not depend on the topology, one can hold that we thereby
sum up all the terms without covariant derivatives in it.

We stress here once again that our analysis is purely local. Of course, there
are always special global effects (Casimir-like effects, influence of
boundaries, closed geodesics etc.) that do not show up in the local expansion
of the heat kernel. The aim of this paper is to study only such situations
where the contribution of these effects is small in comparison with local part,
i.e. the effective action is {\it approximately} given by the integration of
the local formula.

In other words the problem is  the following. One has to obtain a local
covariant function of the invariants of the curvature $\Omega(t)$ (2.9) that
would describe adequately the low-energy limit of the trace of the heat
kernel and that would, being expanded in curvatures, reproduce all terms
without covariant derivatives in the asymptotic expansion of heat kernel,
i.e. the HMDS-coefficients $a_k$ (2.3). If one finds such an expression, then
one can simply determine the $\zeta$-function (2.5) and, therefore, the
low-energy limit of the effective action (2.7), (2.8).

\bigskip
\bigskip
\leftline{\chaptsf III. SYMMETRIC SPACES}
\bigskip
In this paper we will get the most out of the properties of symmetric spaces.
Let us list below some known ideas, facts and formulae about symmetric spaces
presented in the form that is most convenient for
calculating the heat kernel and the effective action.

First of all, we give some definitions (see Refs. 25 and 26 and Sect. III.D).
A Riemannian locally symmetric space which is simply connected and complete is
globally symmetric space (or, simply,  symmetric space)$^{26}$.
A symmetric space is said to be of {\it compact, noncompact {\rm or} Euclidean
type} if {\it all} sectional curvatures $K(u,v)=R_{abcd}u^av^bu^cv^d$
are positive, negative or zero. A direct product of symmetric spaces of compact
and noncompact types is called {\it semisimple} symmetric space. It is well
known$^{25,26}$ that a generic complete simply connected Riemannian symmetric
space is a direct product of a flat space and a semisimple symmetric space (see
also Sect. III.D). Although in the Sect. IY we will need actually only
symmetric spaces of compact type the whole exposition of the Sect. III is valid
for a more general case of semisimple symmetric spaces.

So, what are the direct consequences of the condition of covariant constancy
of the curvature (2.1)?

\bigskip
\leftline{\subchaptsf A. Geometrical framework}
\bigskip

First of all, to carry out the calculations in the curved space in a
covariant way we need some auxiliary two-point geometric objects, namely the
geodetic interval (or world function) $\s(x,x')$, defined as one half
the square of  the length of the geodesic connecting the points $x$ and $x'$,
the tangent vectors $\s_{\m}(x,x')=\na_{\m}\s(x,x')$ and
$\s_{\m'}(x,x')=\na_{\m'}\s(x,x')$ to this geodesic at the points $x$ and $x'$
respectively and a frame $e^\m_a(x,x')$ which is covariantly constant
(parallel) along the geodesic between points $x$ and $x'$, i.e.
$\s^{\m}\na_{\mu}e^\n_a=0.$  We denote the frame components of the tangent
vector by $\s^a(x,x')=g^{ab}e^{\m'}_a(x')\na_{\m'}\s(x,x')$.

Any tensor $T^{a\cdots}_{b\cdots}$, can be presented then in the form of
covariant Taylor series
$$
T^{a\cdots}_{b\cdots} =
\sum_{n\ge 0}{(-1)^n\over n!}\sigma^{\mu'_1}\cdots
\sigma^{\mu'_n}\left[\nabla_{(\mu_1}\cdots\nabla_{\mu_n)}
T^{\a\cdots}_{\b\cdots}\right](x')e^a_{\a'}\cdots e^{\b'}_b.
							\eqno(3.1)
$$
Therefrom it is clear that the frame components of a {\it covariantly} constant
tensor are {\it simply} constant.

In the case of covariantly constant curvature one can express the mixed second
derivatives of the geodetic interval, i.e. the matrix
$$
\s^a_{\ b}(x,x')=e^a_{\m'}(x')e^{\a}_{b}(x)\na^{\m'}\na_\a\s(x,x'),
 \eqno(3.2)
$$
explicitly in terms of the curvature at a fixed point $x'$.
Introducing a matrix $K=\{K^a_{\ b}(x,x')\}$
$$
K^a_{\ b}= R^a_{\ cbd}\s^c\s^d,
							\eqno(3.3)
$$
one can sum up the Taylor series obtaining a closed form$^4$
$$
\s^a_{\ b} = - \left({\sqrt K\over \sin\sqrt K}\right)^a_{\ b}.
		                          \eqno(3.4)
$$
This expression as well as any other similar expressions below should be
always understood as a {\it power series} in the curvature.

\bigskip
\leftline{\subchaptsf B. Curvature}
\bigskip

Let us consider the Riemann tensor in more detail (we follow here the sects.
3.7-3.10 of the first paper in Ref. 25). The components of the curvature tensor
of {\it any} Riemannian manifold can be always presented in the form
$$
R_{abcd} = \b_{ik}E^i_{\ ab}E^k_{\ cd} \eqno(3.5)
$$
where $E^i_{ab}$, $(i=1,\dots, p; p \le d(d-1)/2)$, is some set of
antisymmetric matrices (2-forms) and $\b_{ik}$ is some symmetric nondegenerate
matrix.

Then define the traceless matrices $D_i=\{D^a_{\ ib}\}$
$$
D^a_{\ ib}=-\b_{ik}E^k_{\ cb}g^{ca}= - D^a_{\ bi} \eqno(3.6)
$$
so that
$$
R^a_{\ bcd}=-D^a_{\ ib}E^i_{\ cd},\qquad
R^{a\ c}_{\ b\ d}=\b^{ik}D^a_{\ ib}D^c_{\ kd}, \eqno(3.7)
$$
$$
R^a_b=-\b^{ik}D^a_{\ ic}D^c_{\ kb},\qquad
R=-\b^{ik}D^a_{\ ic}D^c_{\ ka}=-\b^{ik}{\rm tr}(D_i D_k) \eqno(3.8)
$$
where $\b^{ik}=(\b_{ik})^{-1}$.
Because of the curvature identities we have identically
$$
D^a_{\ j [b} E^j_{\ c d]} = 0. \eqno(3.9)
$$
The matrices $D_i$ are known to be the generators of the {\it holonomy
algebra}, i.e. the Lie algebra of the restricted holonomy group,
 ${\cal H}$ (first paper in Ref. 25, p. 97) of dimension ${\rm dim}{\cal H}=p$
$$
[D_i, D_k] = F^j_{\ ik} D_j,\qquad
{\rm or}\qquad
D^a_{\ i c} D^c_{\ k b} - D^a_{\ k c} D^c_{\ i b} = F^j_{\ i k} D^a_{\ j b}.
\eqno(3.10)
$$
The structure constants $F^j_{\ ik}$ of the holonomy algebra are completely
determined by these commutation relations and satisfy the Jacobi identities
$$
F^i_{\ j [k} F^j_{\ m n]} = 0,\qquad
{\rm or}\qquad
[F_i, F_k] = F^j_{\ ik} F_j, \eqno(3.11)
$$
where $F_i=\{F^k_{\ il}\}$ are the generators of the holonomy algebra in
adjoint representation. Note that the restricted holonomy group $H$ is always
compact, as it is a subgroup of the orthogonal group (in Euclidean case), and
connected.

Now let us rewrite the condition of integrability of the relations (2.1) given
simply by the commutator of covariant derivatives
$$
[\na_\m,\na_\n]R_{\a\b\g\d} =-2\left\{ R_{\m\n\l[\a}R^\l_{\ \b]\g\d} +
R_{\m\n\l[\g}R^\l_{\ \d]\a\b}\right\} = 0 \eqno(3.12)
$$
in terms of introduced quantities. It is not difficult to show that it looks
like
$$
E^i_{\ a c} D^c_{\ b k} -  E^i_{\ b c} D^c_{\ a k}= E^j_{\ a b} F^i_{\ j k}.
\eqno(3.13)
$$
This equation takes place {\it only in symmetric spaces} and is the most
important one. It is this equation that makes a Riemannian manifold the
symmetric space.

{}From the eqs. (3.10) and (3.13) we have now
$$
\b_{ik} F^k_{\ jm} + \b_{mk} F^k_{\ ji} = 0, \qquad
{\rm or}\qquad
F^T_i = -\b F_i \b^{-1},     \eqno(3.14)
$$
that means that the adjoint and coadjoint representations of the restricted
holonomy group are equivalent.

The eq. (3.13) leads also to some identities for the curvature tensor
$$
\eqalignno{
&D^{a}_{\ i[b} R_{c]ade}+D^{a}_{\ i[d} R_{e]abc}= 0, &(3.15)\cr
&R^a_{\ c}D^c_{\ ib}=D^a_{\ ic}R^c_{\ b} &(3.16)\cr}
$$
that means, in particular, that the Ricci tensor matrix commutes with all
matrices $D_i$ and is, therefore, an invariant matrix of the holonomy algebra.

Actually, eq. (3.13) brings into existence a much wider algebra ${\cal G}$ of
dimension ${\rm dim}{\cal G}=D=p+d$, in other words it closes this algebra.
Really, let us introduce new quantities $C^A_{\ BC}=-C^A_{\ CB}$, $(A=1,\dots,
D)$
$$
C^i_{\ ab}=E^i_{\ ab}, \quad C^a_{\ ib}=D^a_{\ ib}, \quad C^i_{\ kl}=F^i_{\
kl}, \eqno(3.17)
$$
$$
C^a_{\ bc}=C^i_{\ ka}=C^a_{\ ik}=0,
$$
forming the matrices $C_A = \{C^B_{\ AC}\} = ( C_a, C_i )$
$$
C_a = \left( \matrix{ 0          & D^b_{\ ai}   \cr
		      E^j_{\ ac} & 0            \cr}\right) ,\qquad
C_i = \left( \matrix{ D^b_{\ ia} & 0            \cr
		      0          & F^j_{\ ik}   \cr}\right) ,
						\eqno(3.18)
$$
and symmetric nondegenerate matrix
$$
\g_{AB} = \left(\matrix{ g_{ab} & 0             \cr
	       0                & \b_{ik}        \cr}\right). \eqno(3.19)
$$
Then one can show, first, that as a consequence of the identities (3.9)-(3.13)
the quantities $C^A_{\ CB}$ satisfy the Jacobi identities
$$
C^E_{\ D[A}C^D_{\ BC]}=0,\qquad
{\rm or}\qquad
[C_A, C_B]=C^C_{\ AB}C_C               \eqno(3.20)
$$
and are, therefore, the structure constants of some Lie algebra ${\cal G}$, the
matrices $C_A$ being then the generators of this algebra in adjoint
representation. More precisely, the commutation relations have the form
$$
[C_a, C_b] = E^i_{\ a b} C_i, \qquad
[C_a, C_i] = D^b_{\  a i} C_b, \qquad
[C_i, C_k] = F^j_{\ i k} C_j. \eqno(3.21)
$$
And, second, using the definition of $D$-matrices and the eq. (3.14) one can
show that the structure constants satisfy also the identity
$$
\g_{AB} C^B_{\ CD} + \g_{DB} C^B_{\ CA} = 0,\qquad
{\rm or}\qquad
C^T_A = -\g C_A \g^{-1}, \eqno(3.22)
$$
meaning the equivalence of the adjoint and coadjoint representations of the
algebra ${\cal G}$.

In other words, the Jacobi identities (3.22) are equivalent to the
identities (3.12) that the curvature must satisfy in the symmetric space.
This means that the set of structure constants $C^A_{\ BC}$, satisfying the
Jacobi identities, determines the curvature tensor of symmetric space $R^a_{\
bcd}$. Vice versa the structure of the algebra ${\cal G}$ is completely
determined by the curvature tensor of symmetric space at a fixed point $x'$.

Now consider the curvature of background connection ${\cal R}_{ab}$. One can
show analogously to (3.12) that because of the integrability conditions of the
eq. (2.1)
$$
[\na_\m,\na_\n]{\cal R}_{\a\b} = [{\cal R}_{\m\n},{\cal R}_{\a\b}] -
2R_{\m\n\l[\a}{\cal R}^\l_{\ \b]} = 0 \eqno(3.23)
$$
the curvature of background connection ${\cal R}_{ab}$ in {\it semisimple}
symmetric spaces must have the form
$$
{\cal R}_{ab}={\cal R}_i E^i_{\ ab}, \eqno(3.24)
$$
where $E^i_{\ ab}$ are the same 2-forms and ${\cal R}_i$ are some matrices
forming a representation of the holonomy algebra
$$
[{\cal R}_i, {\cal R}_k]= F^j_{\ ik}{\cal R}_j. \eqno(3.25)
$$
In a generic symmetric space with a flat subspace there are additional Abelian
contributions to the curvature ${\cal R}_{ab}$ (3.24) corresponding to the flat
directions.

Finally, from (2.1) it follows that the potential term should commute with the
curvature ${\cal R}_{\m\n}$
$$
[\na_\m,\na_\n]Q =[{\cal R}_{\m\n}, Q] = 0 \eqno(3.26)
$$
and, therefore, with all the matrices ${\cal R}_i$
$$
[{\cal R}_i, Q]=0. \eqno(3.27)
$$


\bigskip
\bigskip
\leftline{\subchaptsf C. Isometries}
\bigskip

On the covariantly constant background (2.1), i.e. in symmetric spaces, one can
easily solve the Killing
equations
$$
\h L_\xi g_{\m\n} = 2 \na_{(\m}\xi_{\n)} = 0,
							\eqno(3.28)
$$
where $\h L_\xi$ means the Lie derivative.
Indeed, by differentiating the equation
$$
\h L_\xi \Gamma^\l_{\m\n} = \na_{(\m}\na_{\n)}\xi^\l
	+ R^{\l}_{\ (\m|\a|\n)}\xi^\a = 0 ,              \eqno(3.29)
$$
having in mind $\na R = 0$, and symmetrizing the derivatives we get
$$
\eqalignno{
\na_{(\m_1}\cdots\na_{\m_{2n})}\xi^\l &= (-1)^nR^\l_{\ (\m_1|\a_1|\m_2}
R^{\a_1}_{\ \ \m_3|\a_2|\m_4}\cdots R^{\a_{n-1}}_{\ \
\m_{2n-1}|\a_{n}|\m_{2n})}\xi^{\a_{n}}     ,             &(3.30)\cr
\na_{(\m_1}\cdots\na_{\m_{2n+1})}\xi^\l &= (-1)^nR^\l_{\ (\m_1|\a_1|\m_2}
R^{\a_1}_{\ \ \m_3|\a_2|\m_4}\cdots R^{\a_{n-1}}_{\ \
\m_{2n-1}|\a_{n}|\m_{2n}}\na_{\m_{2n+1})}\xi^{\a_{n}}  . &(3.31)\cr}
$$
Thereby we have found all the coefficients of the covariant Taylor series
(3.1) for the Killing vectors of symmetric spaces. Moreover, one can now
sum it up obtaining a closed form
$$
\xi^\m (x) = e^\m_{\ a}\left\{(\cos\sqrt K)^a_{\ b}
\xi^b(x') - \left({\sin\sqrt K\over\sqrt K}\right)^a_{\ b}
\s^{c}\xi^b_{\ ;c}(x')\right\},  \eqno(3.32)
$$
where $\xi^b_{\ ;c}=\xi^\m_{\ ;\n}e^b_\m e^\n_c$.

Therefore, all Killing vectors at any point $x$ are determined in terms  of
initial values of the vectors themselves  $\xi^{b}(x')$ and their
first derivatives $\xi^{b}_{\ ;c}(x')$ at a fixed point $x'$.
The set of all Killing vectors $\hat\h G = \{\xi_{\hat A}\}, (\hat
A=1,\dots\hat D), \dim \hat\h G=\hat D$,
can be split in two essentially different sets: $\h M = \{P_a\}, \dim \h M=d$,
with $P_a$ defined by
$$
P_{\ a}^{\m}(x) = e^\m_{\ b}\left(\cos\sqrt K\right)^{b}_{\ c}P^c_{\ a}(x')
                                \eqno(3.33)
$$
and $\hat\h H = \{L_{\hat i}\}, (\hat i=1,\dots,\hat p), i.e. \dim \hat\h
H=\hat p=\hat D-d $, where
$$
L^\m_{\ \hat i}(x) = - e^\m_{\ b}
	\left({\sin\sqrt K\over \sqrt K}\right)^{b}_{\ a} \s^c
  L^a_{\ \hat i;c}(x'),                        \eqno(3.34)
$$
according to the values of their initial parameters
$$
P^\m_a\Big\vert_{x=x'}\ne 0, \qquad L^\m_{\hat i}\Big\vert_{x=x'}=0.
\eqno(3.35)
$$
Note, that for a general symmetric space $\hat p\ne p$ and, hence, $\hat D\ne
D$!

The Killing vector fields $\xi_{\hat A}=\xi^\m_{\hat A}\na_\m $ (or
$P_a=P^\mu_a \na_\mu$ and $L_{\hat i}=L^\mu_{\ \hat i}\na_\mu$) (acting on
scalar fields) form the Lie {\it algebra of isometries}, $\hat \h G$
$$
[\xi_{\hat A},\xi_{\hat B}]=\hat C^{\hat C}_{\ \hat A\hat B}\xi_{\hat C},
\eqno(3.35a)
$$
or, more explicitly,
$$
[P_a, P_b] = \hat E^{\hat i}_{\ a b} L_{\hat i} , \qquad
[P_a, L_{\hat i}] = \hat D^b_{\  a \hat i} P_b ,
$$
$$
[L_{\hat i}, L_{\hat k}] = \hat F^{\hat j}_{\ \hat i \hat k} L_{\hat j} ,
\eqno(3.35b)
$$
where $\{\hat C^{\hat C}_{\ \hat A\hat B}\}=\{\hat E^{\hat i}_{\ a b}, \hat
D^b_{\  a \hat i}, \hat F^{\hat j}_{\ \hat i \hat k} \}$ are the structure
constants of the algebra of isometries. One sees now that the generators
$L_{\hat i}$ vanishing at the point $x'$ form a subalgebra  (3.35b) of the
algebra of isometries $\hat\h G$ (3.35a) called the {\it isotropy} algebra,
$\hat\h H$.

In fact, all odd symmetrized derivatives of $P^\m_a$ and all even symmetrized
derivatives of $L^\m_{\hat i}$ as well as $L^\m_{\hat i}$ themselves vanish at
the point $x'$
$$
\eqalignno{
&\na_{\n} P_{\ a}^{\m}\Big\vert_{x=x'} =
\na_{(\m_1}\cdots\na_{\m_{2n+1})}P_{\ a}^{\m}\Big\vert_{x=x'} = 0,
							&(3.36)\cr
& L^\m_{\ \hat i}\Big\vert_{x=x'} =
\na_{(\m_1}\cdots\na_{\m_{2n})}L^\m_{\ \hat i}\Big\vert_{x=x'} = 0.
							&(3.37)\cr}
$$
All the parameters $P^b_{\ a}(x')$ are independent and, therefore, there are
exactly $d$ such parameters. The maximal number of the parameters $L^b_{\ \hat
i;c}$ is $d(d-1)/2$, since they are antisymmetric, in other words dim $\hat \h
H\le d(d-1)/2$. However,  they are not independent.
This can be seen immediately if one recalls that the equation
$$
\h L_{L_{\hat i}} R_{\a\b\g\d} = 2 \{ L^\s_{\ \hat i;[\g} R_{\d]\s\b\a} +
	L^\s_{\ \hat i;[\a} R_{\b]\s\d\g} \} = 0
							\eqno(3.38)
$$
holds in symmetric spaces. This equation is, actually, the integrability
condition for  Killing equations (3.26). It imposes strict constraints on the
possible initial parameters $L^{b}_{\ \hat i;c}(x')$.
One can show that for the {\it semisimple} symmetric spaces
the number of independent parameters $L^{b}_{\ \hat i;c}(x')$ is equal to $p$,
i.e. the dimension of the isotropy algebra $\hat\h H$ (3.35b) is equal to the
dimension of the holonomy algebra $\h H$ (3.10),
$$
\hat p\equiv{\rm dim} \hat\h H=\dim \h H\equiv p.
$$
Therefore, the dimension of the algebra of isometries $\hat\h G$, i.e. the
total number of the Killing vectors, in semisimple symmetric spaces is equal to
the dimension of the algebra $\h G$ (3.20) defined in previous Sect. III.B
$$
\hat D\equiv{\rm dim} \hat\h G=\dim \h G\equiv D.
$$
This means that there is no difference between the ordinary latin indices and
the indices with hats. Hence one can omit the hats everywhere. In a symmetric
space of general type having a flat subspace there are additional trivial
Killing vectors corresponding to flat directions. Therefore, in general,
$$
\dim \h H \le \dim\hat\h H \le d(d-1)/2, \qquad
\dim \h G \le \dim\hat\h G \le d(d+1)/2.
$$
The spaces with maximal number of independent isometries, i.e. with
$p=d(d-1)/2$ and $D=d+p=d(d+1)/2$, are the spaces of constant curvature and
only those.

Thus taking into account (3.15) it is evident that one can put
$$
P^a_{\ b}(x')=\d^a_b, \qquad L^a_{\ i;b}(x')=-D^a_{\ ib}. \eqno(3.39)
$$
Therefore, the generators of isometries in semisimple symmetric spaces take the
form
$$
\eqalignno{
&P_{a} = P^\m_{\ a}\na_\m = -
\left(\sqrt K\cot\sqrt K\right)^b_{\ a}\h D_b, &(3.40)\cr
 &L_{i} = L^\m_{\ i}\na_\m = -D^b_{\ ia}\s^a\h D_b,
							&(3.41)\cr }
$$
where
$$
\h D_a = (\s^a_{\ b})^{-1}e^\m_{\ b}\na_\m={\partial\over \partial \s^a}.
							\eqno(3.42)
$$

Moreover, one can show$^{25,26}$ that for semisimple symmetric spaces the
isotropy algebra $\hat\h H$(3.35b) is {\it isomorphic} to the holonomy algebra
$\h H$ (3.10) and the algebra of isometries $\hat\h G$ (3.35a) is isomorphic to
the algebra $\h G$ (3.20) determined by the curvature tensor. Therefore, the
commutation relations (3.35a) and (3.35b) can be rewritten in the form
$$
[\xi_A,\xi_B]=C^C_{\ AB}\xi_C,     \eqno(3.43)
$$
and
$$
[P_a, P_b] = E^i_{\ a b} L_i , \qquad
[P_a, L_i] = D^b_{\  a i} P_b , \qquad
[L_i, L_k] = F^j_{\ i k} L_j, \qquad \eqno(3.44)
$$
with the same structure constants as in (3.20) and (3.21) defined by (3.5),
(3.6), (3.10) and (3.17).
Hence we conclude that the curvature tensor of the {\it semisimple} symmetric
space completely determines the structure of the isotropy algebra and the
algebra of isometries. For a generic symmetric space the curvature determines
the algebra of isometries up to an Abelian ideal. Let us stress once again that
in the case of semisimple symmetric spaces there is no need to distinguish in
notation between the isotropy algebra $\hat\h H$ and the holonomy algebra $\h
H$ and, therefore, between $\hat\h G$ and $\h G$ too.

\bigskip
\bigskip
\leftline{\subchaptsf D. General structure }
\bigskip

As we already noted above the simply connected symmetric space $M$ is
isomorphic to the quotient
space of the group of isometries by the isotropy subgroup $M = \hat G/\hat H
$.$^{25}$
It is, in general, reducible, and has the following general structure$^{25}$
$$
M=M_0\times M_s,
\eqno(3.45a)
$$
$$
M_s=M_+ \times M_-, \eqno(3.45b)
$$
where $M_0$, $M_s$, $M_+$ and $M_-$ are the Euclidean, semisimple, compact and
noncompact components.
The corresponding algebra of isometries is a direct sum of ideals
$$
\hat\h G = \h G_0\oplus\h G_s,     \eqno(3.46)
$$
$$
\h G_s = \h G_+\oplus\h G_-,     \eqno(3.46)
$$
where $\h G_0$ is an Abelian ideal, $\h G_s$ is the semisimple ideal  and $\h
G_+$ and ${\cal G}_-$ are the semi-simple compact and noncompact ones.

There is a remarkable duality relation $*$ between compact and noncompact
objects. For any semisimple algebra of isometries $\h G = \h M + \h H = \{ P_a,
L_k\} $ one defines the dual one according to $\h G^* = i\h M + \h H = \{i P_a,
L_k\}$, the structure constants of the dual algebra being
$$
\{C^{*A}_{\ \ BC}\}=\{E^i_{\ ab}, D^c_{\ dk}, F^j_{\ lm}\}^*=\{ - E^i_{\ ab},
D^c_{\ dk}, F^j_{\ lm}\}.
							\eqno(3.47)
$$
So, the star $*$ only changes the sign of $E^i_{\ ab}$ but does not act
on all other structure constants. This means also that the matrix $\g$ (3.19)
for dual algebra should have the form
$$
\g^{*}_{AB} = \left(\matrix{ g_{ab} & 0             \cr
	       0                & \b_{ik}        \cr}\right)^*
=\left(\matrix{ g_{ab} & 0             \cr
	       0                & -\b_{ik}        \cr}\right) \eqno(3.48)
$$
and, therefore, the curvature of the dual manifold has the opposite sign
$$
R^*_{abcd}=-R_{abcd}. \eqno(3.49)
$$


\bigskip
\bigskip
\leftline{\chaptsf IY. HEAT KERNEL}
\bigskip

It should be noted once more that our analysis in this paper is purely {\it
local}. (See the discussion in Sect. II.) We are looking for a {\it universal
local} function of the curvature, $\Omega(t)$, (2.9) that describes adequately
the low-energy limit of the heat kernel diagonal (up to `global' nonanalytical
effects that are not studied in this paper!). Our minimal requirement is that
this function should reproduce {\it all} the terms without covariant
derivatives of the curvature in the local Schwinger-De Witt asymptotic
expansion of the heat kernel, i.e.
it should give {\it all} the HMDS-coefficients $a_k$ (2.3) for {\it any}
symmetric space.

It is well known that the HMDS-coefficients have a {\it universal} explicit
structure$^5$, i.e. $a_k$ are scalar polynomials of the curvature of the order
$k$ with {\it universal} numerical coefficients that do not depend on the
particular form of the symmetric space, on the dimension etc. It is obvious
that any flat subspaces do not contribute in $a_k$. Moreover, since
HMDS-coefficients $a_k$ are analytic in the curvature it is evident that
to find this universal structure it is sufficient to consider only symmetric
spaces of {\it compact} type with positive curvature.
Using the factorization property of the heat kernel$^5$ and the duality between
compact and noncompact symmetric spaces we can obtain then the results for the
general case by analytical continuation.

That is why below in this paper we consider only the case of symmetric spaces
of {\it compact} type  when the matrices $\b_{ik}$ and  $\g_{AB}$ are {\it
positive} definite. Besides, we restrict ouselves, for simplicity,  to the
scalar operators, i.e. ${\cal R}_{\a\b}=0$. The general case will be
investigated in a future work.

\bigskip
\leftline{\subchaptsf A. Heat kernel operator}
\vglue0pt
\bigskip
\vglue0pt
It is not difficult to show that the metric of the symmetric space can be
presented in the form
$$
g^{\m\n} = \g^{AB}\xi^\m_A\xi^\n_B = g^{ab}P^\m_a P^\n_b
+ \b^{ik}L^\m_i L^\n_k .
							\eqno(4.1)
$$
Indeed, by making use of the eqs. (3.7) and recalling the definition of the
matrix $K$ (3.3) it is easy to obtain (4.1) using the explicit expressions
(3.33), (3.34).

Now having the metric (4.1) we can build the Laplacian for the scalar (${\cal
R}_{\a\b}=0$) case
$$
\sq = g^{\m\n}\na_\m\na_\n = \g^{AB}\xi_A\xi_B ,         \eqno(4.2)
$$
where $\xi_A=\xi^\m_A\na_\m $ and the Killing equation (3.5) has been used.

It is not difficult to show that the Laplacian belongs to the center of the
enveloping algebra, i.e. it commutes with all the generators of the algebra
$$
[\sq, \xi_A] = 0 .                                       \eqno(4.3)
$$
Let us now try to represent the heat kernel in terms of a group average, i.e.
let us find a formula like
$$
\exp{(t\sq)} = \int d k \g^{1/2}\Phi (t|k) \exp(k^A\xi_A) .  \eqno(4.4)
$$
We formulate first the answer in form of a theorem and prove it
below.\hb
{\sc Theorem 1:}\hb {\sl
For any compact $D$-dimensional Lie group generated by $\xi_A$
$$
[\xi_A,\xi_B]=C^C_{\ AB}\xi_C                           \eqno(4.5)
$$
it takes place the operator identity
$$
\eqalignno{
\exp(t\sq) = (4\pi t)^{-D/2} \int d k \g^{1/2}
	&\det\left({\sinh(k^AC_A/2)\over k^AC_A/2}\right)^{1/2} &\cr
	& \times\exp\left\{ -{1\over 4t}k^A\g_{AB}k^B
	+ {1\over 6} R_G t\right\}\exp(k^A\xi_A) ,        &(4.6)\cr}
$$
where $\sq=\g^{AB}\xi_A\xi_B$, $\g^{AB} = (\g_{AB})^{-1}$, $\g=\det\g_{AB}$,
$\g_{AB}$ is a symmetric nondegenerate positive definite matrix connecting the
generators
in adjoint $C_A=(C^B_{\ AC})$ and co-adjoint $C_A^T$ representations
$$
C_A^T = -\g C_A \g^{-1},                                \eqno(4.7)
$$
$R_G$ is the scalar curvature of the group manifold
$$
R_G= -{1\over 4}\g^{AB} C^C_{\ AD}C^D_{\ BC} ,       \eqno(4.8)
$$
and the integration is to be taken over the whole Euclidean
space $\RR^D$.\hb}
{\sl The proof:}\hb
Let us consider the integral
$$
\Psi(t) = \int d k \g^{1/2}\Phi (k,t) \exp(k^A\xi_A),    \eqno(4.9)
$$
where
$$
\eqalignno{
\Phi(t|k) = &(4\pi t)^{-D/2}
		\det\left({\sinh(k^AC_A/2)\over k^AC_A/2}\right)^{1/2}
\exp\left\{ -{1\over 4t}k^A\g_{AB}k^B
	+ {1\over 6} R_G t\right\}.        &(4.10)\cr}
$$
To prove the theorem we have to show that $\Psi(t)=\exp(t\sq)$, in other words,
that it satisfies the operator equation
$$
\partial_t \Psi = \sq \Psi                              \eqno(4.11)
$$
with initial condition
$$
\Psi(t)\Big\vert_{t=0} = 1.                           \eqno(4.12)
$$

First one can show that
$$
\xi_B \exp(k^A\xi_A) = X_B \exp(k^A\xi_A),           \eqno(4.13)
$$
where
$$
 X_A =  X^M_{\ A} (k) {\partial\over\partial k^M}    \eqno(4.14)
 $$
 are the left-invariant vector fields on the group that have in canonical
 coordinates the explicit form
 $$
 X^M_{\ A}(k) = \left({k^AC_A\over \exp(k^AC_A) -1}\right)^M_{\ A}.
							\eqno(4.15)
 $$
 Therefore, from the definition of the Laplacian we have
 $$
 \eqalignno{
 &\sq\exp(k^A\xi_A) = X_2 \exp(k^A\xi_A) ,               &(4.16)\cr
 &X_2 = \g^{AB} X_A X_B .                                &(4.17)\cr}
 $$
 Then, introducing the metric on the group manifold
 $$
 G_{MN} = \g_{AB} X^{-1 A}_{\ \ \ \ M} X^{-1 B}_{\ \ \ \ N} \eqno(4.18)
 $$
 and its determinant
 $$
 G=\det G_{MN}=\g\det X^{-2}
 = \g \det\left({\sinh(k^AC_A/2)\over k^AC_A/2}\right)^2,    \eqno(4.19)
 $$
 one can obtain the transposition relation
 $$
 \left(G^{1/2}X_2 G^{-1/2}\right)^{T} = X_2.           \eqno(4.20)
 $$
 Now, making use of (4.9), (4.16) and (4.20) and integrating by parts
 we obtain
 $$
 \sq \Psi(t) =\int d k \g^{1/2}\exp(k^A\xi_A)
 \left(G^{1/2}X_2 G^{-1/2}\Phi \right) .            \eqno(4.21)
$$
On the other hand, one has from (4.9)
$$
\partial_t \Psi(t) = \int d k \g^{1/2} \partial_t \Phi \exp(k^A \xi_A).
							\eqno(4.22)
$$
Thus to prove (4.11) we have to show that
$$
\partial_t \Phi = G^{1/2}X_2 G^{-1/2}\Phi .             \eqno(4.23)
$$
Substituting the explicit expression for $\Phi$
$$
\Phi (t|k) = \g^{-1/4} G^{1/4}(k)(4\pi t)^{-D/2}
\exp\left\{-{1\over 4t} k^A\g_{AB} k^B
+ {1\over 6} R_G t\right\}                          \eqno(4.24)
$$
and using the relations
$$
X_2 G^{-1/4} = {1\over 6} R_G G^{-1/4}              \eqno(4.25)
$$
and
$$
k^A{\partial\over \partial k^A} G^{-1/4} = {1\over 2} (D - {\rm tr} X)
G^{-1/4},                                               \eqno(4.26)
$$
where
$$
{\rm tr} X=X^A_A={\rm tr}\left({k^AC_A} \coth\left({k^AC_A}\right) \right),
							\eqno(4.27)
$$
that hold on the group manifold,  we convince ourselves that the eq. (4.23)
is correct.
Thereby it is shown that $\Psi(t)$ really satisfies the eq. (4.11).

Further, from (4.10) it follows immediately
$$
\Phi(t|k)\Big\vert_{t=0} = \g^{-1/2}\d(k)                   \eqno(4.28)
$$
and, therefore, the initial condition (4.12). Thus we found
$\Psi(t)=\exp(t\sq)$ that proves the theorem.

\bigskip
\bigskip
\leftline{\subchaptsf B. Heat kernel diagonal}
\bigskip
So, we have found a very nontrivial representation (4.6) that holds on any
compact Lie group.
How can we proceed now with this useful theorem?

First, we can express the scalar curvature of the
group manifold in terms of the scalar curvature of the symmetric space $R$
and that of the isotropy subgroup $R_H $
$$
R_G = -{1\over 4} \g^{AB} C^C_{\ AD}C^D_{\ BC}
= {3\over 4} R + R_H ,                            \eqno(4.29)
$$
where
$$
R_H = -{1\over 4} \b^{ik} F^m_{\ \ il}F^l_{\ km}. \eqno(4.30)
$$

The representation (4.6) is valid for any generators $\xi_A$, satisfying the
commutation relations (4.5), and so it is also valid for the  infinitesimal
isometries (3.40), (3.41) of the symmetric space. In this case $\sq$ is the
usual  Laplacian and $\exp(t\sq)$ is the heat kernel operator.

For further use it is convenient to rewrite the integral (4.6)
splitting the integration variables $k^A = (q^a, \om^i)$ in the form
$$
\eqalignno{
\exp(t\sq) = &(4\pi t)^{-D/2} \int dq\,d\om \eta^{1/2}\b^{1/2}
\det\left({\sinh((q^a C_a + \om^i C_i)/2)\over (q^a C_a + \om^i C_i)/2}
\right)^{1/2} &\cr
& \times\exp\left\{ -{1\over 4 t}(q^a g_{ab}q^b + \om^i\b_{ik}\om^k)
+ \left({1\over 8} R + {1\over 6} R_H \right) t \right\}
\exp\left(q^a P_a + \om^i L_i\right) ,  &\cr
&                 &(4.31)\cr}
$$
where$\b=\det \b_{ik}$, $\eta=\det g_{ab}$.
To get the heat kernel explicitly in coordinate representation we have to act
with the heat kernel operator $\exp(t\sq)$ on the delta-function on $M$
$$
\exp(t\sq)(x,x') = \exp(t\sq)\d(x,x')
=\int dq\,d\om \eta^{1/2}\b^{1/2} \Phi(t|q,\om)
\exp\left(q^a P_a + \om^i L_i\right)\d(x,x').              \eqno(4.32)
$$
To learn how the operator $\exp(k^A\xi_A)$ acts on a scalar function $f(x)$
let us introduce a new function
$$
\phi(s,k,x) = \exp(s k^A \xi_A) f(x) .                  \eqno(4.33)
$$
This function satisfies the first order differential equation
$$
\partial_s\phi = k^A\xi_A\phi = k^A\xi^\m_{\ A}(x)\partial_\m\phi \eqno(4.34)
$$
with the initial condition of the form
$$
\phi\Big\vert_{s=0}=f(x) .                           \eqno(4.35)
$$
It is not difficult to prove that
$$
\phi(s,k,x)=f(x_0(s,k,x)),                          \eqno(4.36)
$$
where $x_0(s,k,x)$ satisfies the equation of characteristics
$$
{d x_0^\m\over ds} = k^A\xi^\m_A (x_0)            \eqno(4.37)
$$
with initial condition
$$
x_0^\m\Big\vert_{s=0} = x^\m .                        \eqno(4.38)
$$
Therefore, we have
$$
\exp\left(k^A\xi_A\right)\d(x,x') = \d(x_0(1,k,x),x').   \eqno(4.39)
$$

Consider now the operator integrals of the form we need
$$
I(x,x') = \int dq\, d\om \eta^{1/2}\b^{1/2}Z(q,\om)
\exp\left(q^a P_a + \om^i L_i\right) \d (x,x'),           \eqno(4.40)
$$
where $Z(q,\om)$ is some analytic function.
Using the eq. (4.39) we have
$$
\exp\left(q^a P_a + \om^i L_i\right)\d(x,x')
= \d(x_0(1,q,\om,x,x'),x')
= \eta^{-1/2}J(\om,x,x')\d(q-\bar q) ,        \eqno(4.41)
$$
where $\bar q=\bar q(\om,x,x')$ is to be determined from the equation
$$
x_0(1, \bar q, \om,x,x')=x'                          \eqno(4.42)
$$
and $J(\om,x,x')$ is the Jacobian computed at $x_0=x'$
$$
J(\om,x,x') = g^{\prime -1/2}\eta^{1/2}
\det\left|{\partial x_0^\m \over \partial q^a}
\right|^{-1}_{q=\bar q, s=1} .                      \eqno(4.43)
$$
So, we can now simply integrate over $q$ in (4.40) to get
$$
I(x,x') = \int d\om \b^{1/2}Z(\bar q(\om,x,x'),\om)J(\om,x,x').
							\eqno(4.44)
$$
If we are interested in coincidence limit then one has to put finally $x=x'$
$$
I(x,x) = \int d\om \b^{1/2}Z(\bar q(\om,x,x),\om) J(\om,x,x).
							\eqno(4.45)
$$
Consider now the equation of characteristics at greater length. Making a
change of variables
$$
x^\m \to \s^a_0 = \s^a(x_0,x')
= e^{a}_{\m'}(x')\s^{\m'}(x_0,x')                \eqno(4.46)
$$
we arrive to the equation of more explicit form
$$
{d \s_0^a\over ds} = - \left(\sqrt {K(\s_0)}\cot \sqrt
 {K(\s_0)}\right)^a_{\ b}q^b - \om^iD^a_{\ ib}\s^b_0 . \eqno(4.47)
$$

Let $\s_0^a=\s_0^a(s,q,\om,\s^b)$ be the solution of the equation (4.47).
Then $\bar q$ is to be determined from an equation like (4.42)
$$
\s_0^a(1,\bar q,\om,\s^b)=0                              \eqno(4.48)
$$
and
$$
J(\om,x,x') = \det\left|-{\partial \s_0^a \over \partial
q^b}\right|^{-1}_{q=\bar q, s=1} ,                         \eqno(4.49)
$$
where it has been taken into account $\det(e^{\m'}_a)=g^{\prime -1/2}
\eta^{1/2}$.

Therefore, we have to find the solution to the equation (4.47) near the zero,
i.e. assuming $\s_0^a$  to be small.
Moreover, we consider mostly the case when the points $x$ and $x'$ are close
to each other that means that $\s^a$ is small too. The equation (4.47) near
the point $\s_0^a=0$ looks like
$$
{d \s_0^a\over ds} = - q^a                             \eqno(4.50)
$$
meaning that the momentums $q^a$ are of the same small order.

More precisely, we assume
$$
\s_0^a \sim \s^b \sim q^c \sim \eps \ll 1               \eqno(4.51)
$$
and look for a solution of the eq. (4.47) in form of a power series in
$\eps$, i.e. in form of a Taylor series in $\s^a$ and $q^a$.

In this way one simply obtains up to quadratic terms
$$
\s_0^a(s,q,\om,x,x') = (\exp(-s \om^iD_i))^a_{\ b}\s^b
+ \left({\exp(-s \om^iD_i)-1\over \om^iD_i}\right)^a_{\ b}q^b + O(\eps^2)
							\eqno(4.52)
$$

With the same accuracy the solution of the eq. (4.48) is
$$
\bar q^a =
\left({\om^iD_i\exp(-s \om^iD_i) \over 1-\exp(-s \om^iD_i)}\right)^a_{\ b}\s^b
+ O(\s^{a\, 2}).                                     \eqno(4.53)
 $$

Further, one finds from (4.52)
$$
\det\left|-{\partial \s_0^a \over \partial q^b}
\right|_{q=\bar q, s=1}
= \det\left({\sinh(\om^iD_i/2)\over \om^iD_i/2}\right) + O(\s^a) \eqno(4.54)
$$
and so, from (4.49)
$$
J(\om,x,x')=\det\left({\sinh(\om^iD_i/2)\over \om^iD_i/2}\right)^{-1} +
O(\s^a).
								\eqno(4.55)
$$

Substituting (4.53) and (4.55) in (4.44) and expanding $Z(\bar q,\om)$
we can calculate the integral (4.40) for near points $x$ and $x'$ in form
of an expansion in $\s^a(x,x')$.

Therefore, we have found, in particular, a useful exact result for coincidence
limit (4.45).\hb
{\sc Lemma 2:}\hb
{\sl
For an analytical function $Z(q,\om)$ there holds
$$
\eqalignno{
I(x,x) = & \int dq\, d\om \eta^{1/2}\b^{1/2}Z(q,\om)
\exp\left(q^a P_a + \om^i L_i\right) \d (x,x')\Big\vert_{x=x'} &\cr
       = & \int d\om \b^{1/2}Z(0,\om)
	\det\left({\sinh(\om^iD_i/2)\over \om^iD_i/2}\right)^{-1} &(4.56)\cr}
$$
with the operators $P_a$ and $L_i$ given by (3.40) and (3.41).}


Using the obtained results (4.53), (4.55) and (4.56) and substituting
the explicit form of our
integral (4.31)  we get the heat kernel in coordinate representation
$$
\eqalignno{
\exp(t\sq)(x,x') = &(4\pi t)^{-D/2}\int d\om \b^{1/2}_{H}
\det\left({\sinh(\om^iC_i/2)\over \om^iC_i/2}\right)^{1/2}
\det\left({\sinh(\om^iD_i/2)\over \om^iD_i/2}\right)^{-1} &\cr
& \times\exp\left\{ - {1\over 4 t}(\om^i\b_{ik}\om^k
+ \s^a g_{ac}B^c_{\ b}(\om)\s^b)
+ \left({1\over 8} R + {1\over 6} R_H \right) t \right\}
+ O(\s^a),    &\cr
&                                           &(4.57)\cr}
$$
where $B(\om) = \{B^a_{\ b}(\om)\}$ is a matrix of the form
$$
B(\om) = \left({\sinh(\om^iD_i/2) \over \om^iD_i/2}\right)^{-2}.
							\eqno(4.58)
$$

Now, from (3.18) it is not difficult to find that
$$
\det\left({\sinh(\om^iC_i/2)\over \om^iC_i/2}\right) =
\det\left({\sinh(\om^iD_i/2)\over \om^iD_i/2}\right)
\det\left({\sinh(\om^iF_i/2)\over \om^iF_i/2}\right).
							\eqno(4.59)
$$
Therefore, the final result after taking into account (4.59) looks like
$$
\eqalignno{
\exp(t\sq)(x,x') = &(4\pi t)^{-D/2}\int d\om \b^{1/2}
\det\left({\sinh(\om^iF_i/2)\over \om^iF_i/2}\right)^{1/2}
\det\left({\sinh(\om^iD_i/2)\over \om^iD_i/2}\right)^{-1/2} &\cr
&\times\exp\left\{ - {1\over 4 t}(\om^i\b_{ik}\om^k
+ \s^a g_{ac}B^c_{b}(\om)\s^b) + \left({1\over 8} R
+ {1\over 6} R_H \right) t \right\} + O(\s^a). &\cr
&           &(4.60)\cr}
$$
The coincidence limit of this heat kernel is then simply derived by putting
$x=x'$, i.e. $\s^a=0$,
$$
\eqalignno{
\exp(t\sq)(x,x) = &(4\pi t)^{-D/2}\int d\om \b^{1/2}
\det\left({\sinh(\om^iF_i/2)\over \om^iF_i/2}\right)^{1/2}
\det\left({\sinh(\om^iD_i/2)\over \om^iD_i/2}\right)^{-1/2} &\cr
&\times\exp\left\{ - {1\over 4 t}\om^i\b_{ik}\om^k
+ \left({1\over 8} R
+ {1\over 6} R_H \right) t \right\}                     &(4.61)\cr}
$$

Note, that this formula is exact (up to possible nonanalytic topological
contributions, see the discussion in sect. II).
This gives a nontrivial example how the heat kernel can be constructed using
only the algebraic properties of the isometries of the symmetric space.

One can derive an alternative nontrivial {\it formal} representation of this
result. Substituting the equation
$$
(4\pi t)^{-p/2}\b^{1/2}\exp\left(-{1\over 4t} \om^i\b_{ik}\om^k\right) =
(2\pi)^{-p}\int dp \exp \left(ip_k\om^k -tp_k\b^{kn}p_n\right) \eqno(4.62)
$$
into the integral (4.61) and integrating over $\om$ we obtain
$$
\eqalignno{
\exp(t&\sq)(x,x) = (4\pi t)^{-d/2}
\exp\left\{t\left({1\over 8} R + {1\over 6} R_H \right)\right\} &\cr
&\times\int dp  \exp\left(- t p_n\b^{nk}p_k\right)
\det\left({\sinh(-i\partial^kF_k/2)\over -i\partial^kF_k/2}\right)^{1/2}
\det\left({\sinh(-i\partial^kD_k/2)\over -i\partial^kD_k/2}\right)^{-1/2}\d(p),
&\cr
& &(4.63)\cr}
$$
where $\partial^k=\partial/\partial p_k$.
Therefrom integrating by parts and changing the integration variables $p_k \to
i t^{-1/2} p_k $ we get finally an expression without any integration
$$
\eqalignno{
\exp(t&\sq)(x,x) = (4\pi t)^{-d/2}\exp\left(t\left({1\over 8} R
+ {1\over 6} R_H \right)\right) &\cr
&\times\det\left({\sinh(\sqrt t \partial^kF_k/2)\over \sqrt t
\partial^kF_k/2}\right)^{1/2}
\det\left({\sinh(\sqrt t \partial^kD_k/2)\over \sqrt t
\partial^kD_k/2}\right)^{-1/2}
\exp\left(p_n\b^{nk}p_k\right)\Bigg\vert_{p=0}. &\cr
&       &(4.64)\cr}
$$
This formal solution should be understood as a power series in the derivatives
$\partial^i$ that is well defined and determines the heat kernel asymptotic
expansion at $t\to 0$.

\bigskip
\bigskip
\leftline{\subchaptsf C. Heat kernel asymptotics }
\bigskip

Using obtained result one can get easily the explicit form of the generating
function for HMDS-coefficients (2.9)
$$
\eqalignno{
\Omega(t|x,x) = &(4\pi t)^{-p/2}\int d\om \b^{1/2}
\det\left({\sinh(\om^iF_i/2)\over \om^iF_i/2}\right)^{1/2}
\det\left({\sinh(\om^iD_i/2)\over \om^iD_i/2}\right)^{-1/2} &\cr
&\times\exp\left\{ - {1\over 4 t}\om^i\b_{ik}\om^k
+ \left({1\over 8} R
+ {1\over 6} R_H \right) t \right\}                     &(4.65)\cr}
$$

This formula can be used now to generate {\it all} HMDS-coefficients
$a_k$ for {\it any} symmetric space, i.e. for {\it any space with covariantly
constant curvature}, simply by expanding it in a power series in $t$.

Changing the integration variables $\om \to \sqrt t \om$ and introducing a
Gaussian averaging over $\om$
$$
<f(\om)> = (4\pi)^{-p/2}\int d \om \b^{1/2}
\exp\left(-{1\over 4}\om^i\b_{ik}\om^k \right) f(\om)
							\eqno(4.66)
$$
we get
$$
\eqalignno{
\Omega(t|x,x) = &
\exp\left\{\left({1\over 8} R + {1\over 6} R_H \right) t \right\}
\Bigg<\det\left({\sinh(\sqrt t \om^iF_i/2)\over \sqrt t
\om^iF_i/2}\right)^{1/2}
\det\left({\sinh(\sqrt t \om^iD_i/2)\over \sqrt t
\om^iD_i/2}\right)^{-1/2}\Bigg> &\cr
& &(4.67)\cr}
$$

Using the standard Gaussian averages
$$
\eqalignno{
&<1> = 1 \qquad,\qquad <\om^i> = 0 \qquad,\qquad
 <\om^i\om^k> = {1\over 2}\b^{ik}                     &\cr
&<\om^i_1\cdots \om^{i_{2n+1}}> = 0 ,                   &(4.68)\cr
&<\om^{i_1}\cdots \om^{i_{2n}}> = {(2n)!\over 2^{2n}n!}\b^{(i_1 i_2}\cdots
\b^{i_{2n-1}i_{2n})}                                  &\cr}
$$
one can obtain now all HMDS-coefficients in terms of various foldings of the
quantities $D^a_{\ ib}$ and $F^j_{\ ik}$ with  the help of matrix $\b^{ik}$.
All these quantities are curvature invariants and can be expressed directly in
terms of Riemann tensor. Thereby one finds {\it all covariantly constant terms
in {\it all} HMDS-coefficients} in manifestly covariant way. We are going to
obtain the explicit formulae in a further work.

\bigskip
\bigskip
\leftline{\chaptsf Y. CONCLUDING REMARKS}
\bigskip

In present paper we continued the study of the heat kernel that we conducted
in our papers (Ref. 4,10,15,21).
Here we have discussed some ideas connected with the point that was
left aside in previous papers, namely, the problem of calculating the
low-energy limit of the effective action in quantum gravity.
We have analyzed in detail  the status of the low-energy limit in quantum
gravity and stressed the central role  playing
by the Lie group of isometries that naturally appears when
generalizing consistently the low-energy limit to curved space.

We have proposed a promising, to our mind, approach for  calculating the
low-energy heat kernel and realized, thereby, the idea of partial
summation of the terms without covariant derivatives in local
asymptotic expansion for computing the effective action that was
suggested in Ref. 2,4.

Of course, there are left many unsolved problems. First of all, one has to
obtain {\it explicitly} the covariantly constant terms in HMDS-coefficients.
This would be the opposite case to the high-derivative approximation$^{15,16}$
and can be of certain interest in mathematical physics. Then, we still do not
know how to calculate the low-energy heat kernel in general case of covariantly
constant curvatures, i.e. when all background curvatures ( $\Re=\{R_{\m\n\a\b},
{\cal R}_{\m\n}, Q \}$) are present.
 Besides, it is not perfectly clear how to do the
analytical continuation of Euclidean  low-energy effective action to the
space of Lorentzian signature for obtaining physical results.

Let us make a final remark concerning the relation of our work to that of
previous authors who seems to treat almost the same problem (see the review
paper of Camporesi in Ref. 22 and references therein and Ref. 27). What we have
been trying to do in present paper is rather different from what the other
authors did. These are the {\it global} topological problems and effects that
are of {\it prime} interest in those papers. The authors of those papers make
use of the techniques of geometric analysis on homogenous spaces with emphasis
on {\it exact} results. That is why only very special concrete examples of
symmetric spaces (group manifolds, spheres, rank-one symmetric spaces,
split-rank symmetric spaces etc.) which allow to obtain closed formulas were
considered. The results obtained in this way are presented in terms of the root
vectors and their multiplicities. The complexity of the method depends
critically on the rank of symmetric space. As far as we know the explicit
results for the heat kernel are obt
ained for rank-one and for some rank-two symmetric spaces.

We are interested, in contrary, first of all in {\it local} effects of strongly
curved {\it approximately homogeneous} manifolds. Therefore, our approach is
thought of only as a {\it framework for a perturbation theory} in
non-homogeneity. In typical physical problems we need rather {\it general
approximation scheme} instead of exact exceptional results. The point is we
need the effective action as a functional of a {\it generic} metric which could
be varied to obtain the physical currents.

There is, of course, the difficult question, whether the global effects might
be neglected in comparison with local ones. This question is {\it open}. We can
only say that if it is the case, i.e. {\it if the local effects are dominant},
then the heat kernel is given by explicit covariant formulas obtained in sect.
IY.

\bigskip
\bigskip
\leftline{\chaptsf Y. NOTE ADDED}
\bigskip

After this paper was completed we became aware of the similar results on heat
kernel in symmetric spaces by Fegan (Ref. 27).
Though they were obtained in completely different  rather geometrical setting
incorporating the nontrivial global topology, one would, due to intrinsic
locality of the heat kernel expansion, expect that the two expressions, i.e
ours and that of Ref. 27, should coincide under an appropriate representation
of the special functions obtained in Ref. 27.

Another comment concerns the meaning of the effective potential. If the
symmetry in question is that of Euclidean space (that is not the case, in
general!) our expansion should reduce to the quasi-local expansion of Brown and
Duff$^{28}$, which was extended to curved spaces by Hu and O'Connor.$^{29}$
Therefore, one might consider our work as an extension of the quasi-local
expansion to a symmetric space and quasi-homogenous setting.

%
%

\bigskip
\bigskip
\leftline{\chaptsf ACKNOWLEDGEMENTS }
\bigskip

I would like to thank G. A. Vilkovisky for many helpful discussions and R.
Schimming and J. Eichhorn for their hospitality at the University of
Greifswald. I am
also grateful to P. B. Gilkey, H. Osborn, S. Fulling, D. M. Mc Avity,  T.
Osborn,
S. Odintsov and K.  Kirsten for correspondence.
This work was supported, in part, by the Alexander von Humboldt Foundation,
by a Soros Humanitarian Foundations Grant
awarded by the American Physical Society and by an Award
  through the International Science Foundation's Emergency Grant
  competition.

\bigskip
\bigskip

\item{$^1$}  B. S. De Witt,
		{\it Dynamical theory of groups and fields}, (Gordon and
		Breach,
		New York, 1965);
		in: {\it Relativity, groups and topology II}, ed. by B. S.  De
		Witt
		and R. Stora, (North Holland, Amsterdam, 1984) p. 393.
\item{$^2$} G. A. Vilkovisky,
		in: {\it Quantum theory of gravity}, ed.  S. Christensen
		(Hilger,
		Bristol, 1983) p. 169.
\item{$^3$} A. O. Barvinsky and G. A. Vilkovisky,
		Phys. Rep. {\bf C 119}, 1 (1985).
\item{$^4$} I. G. Avramidi,
		Nucl. Phys. {\bf B 355}, 712 (1991).
\item{$^5$} P. B. Gilkey,
		{\it Invariance theory, the heat equation and the  Atiyah -
		Singer
		index theorem}, (Publish or Perish, Wilmington, DE, USA, 1984).
\item{$^6$} J. Hadamard,
		{\it Lectures on Cauchy's Problem}, in: {\it Linear Partial
		Differential
		Equations}, (Yale U. P., New Haven, 1923);
\item{} S. Minakshisundaram and A. Pleijel,
		Can. J. Math. {\bf 1}, 242 (1949);
\item{} R. T. Seeley,
		Proc. Symp. Pure Math. {\bf 10}, 288 (1967);
\item{} H. Widom,
		Bull. Sci. Math. {\bf 104}, 19 (1980);
\item{} R. Schimming,
		Beitr. Anal. {\bf 15}, 77 (1981);
		Math. Nachr. {\bf 148}, 145 (1990).
\item{$^7$} S. A. Fulling and G. Kennedy,
		Trans. Am. Math. Soc. {\bf 310}, 583 (1988).
\item{$^8$}  V. P. Gusynin,
		Phys. Lett. {\bf B 255}, 233 (1989).
\item{$^9$} P. B. Gilkey,
		J. Diff. Geom. {\bf 10}, 601 (1975).
\item{$^{10}$} I. G. Avramidi,
		Teor. Mat. Fiz. {\bf 79}, 219 (1989);
		Phys. Lett. {\bf B 238}, 92 (1990).
\item{$^{11}$} P. Amsterdamski, A. L. Berkin and D. J. O'Connor,
		Class. Quantum  Grav. {\bf 6}, 1981 (1989).
\item{$^{12}$} T. P. Branson  and P. B. Gilkey,
		Comm. Part. Diff. Eq. {\bf 15}, 245 (1990);
\item{}     N. Blazic, N. Bokan and P. B. Gilkey,
		Indian J. Pure Appl. Math. {\bf 23}, 103 (1992).
\item{$^{13}$} G. Cognola, L. Vanzo and S. Zerbini,
		Phys. Lett. {\bf B 241}, 381 (1990);
\item{}     D. M. Mc Avity and H. Osborn,
		Class. Quantum Grav. {\bf 8}, 603 (1991);
		Class. Quantum Grav. {\bf 8}, 1445 (1991);
		Nucl. Phys. {\bf B 394}, 728 (1993);
\item{}     A. Dettki and A. Wipf,
		Nucl. Phys. {\bf B 377}, 252 (1992);
\item{}     I. G. Avramidi,
		Yad. Fiz. {\bf 56}, 245 (1993).
\item{$^{14}$} P. B. Gilkey,
		{\it Functorality and heat equation asymptotics}, in: Colloquia
		Mathematica Societatis Janos Bolyai, 56. Differential
		Geometry, (Eger (Hungary), 1989), (North-Holland, Amsterdam,
		1992), p.  285;
\item{} R. Schimming,
		{\it Calculation of the heat kernel coefficients}, in {\it Analysis, Geometry
and Groups: A Riemann Legacy Volume}, edited by H. Srivastava and Th. M.
Rassias, (Hadronic Press, Palm Harbor, 1993), part. II, p. 627.

\item{$^{15}$} I. G. Avramidi,
		Yad. Fiz. {\bf 49}, 1185 (1989);
		Phys. Lett. {\bf B 236}, 443 (1990).
\item{$^{16}$}     T. Branson, P. B. Gilkey and B. \O rsted,
		Proc. Amer. Math. Soc. {\bf 109}, 437 (1990).
\item{$^{17}$} A. O. Barvinsky and G. A. Vilkovisky,
		Nucl. Phys. {\bf B 282}, 163 (1987);
		Nucl. Phys. {\bf B 333}, 471 (1990);
\item{}     G. A. Vilkovisky,
		{\it Heat kernel: recontre entre physiciens et mathematiciens}, in: Proc. of
Strasbourg
		Meeting between physicists and  mathematicians (Publication de
		l' Institut de Recherche Math\'ematique  Avanc\'ee,
		Universit\'e  Louis
		Pasteur, R.C.P. 25, vol.43 (Strasbourg, 1992)), p. 203;

\item{}     A. O. Barvinsky, Yu. V. Gusev, G. A. Vilkovisky and V. V.
Zhytnikov, J. Math. Phys. {\bf 35}, 3525 (1994).
\item{$^{18}$} F. H. Molzahn, T. A. Osborn and S. A. Fulling,
		Ann. Phys. (USA) {\bf 204}, 64 (1990);
\item{} P. B. Gilkey, T. P. Branson and S. A. Fulling,
		J. Math. Phys. {\bf 32}, 2089 (1991);
\item{} T. P. Branson, P. B. Gilkey and A. Pierzchalski,
		Math. Nachr. {\bf 166}, 207 (1994);
\item{} A. O. Barvinsky, T. A. Osborn and Yu. V. Gusev,
J. Math. Phys. {\bf 36}, 30 (1995).
\item{$^{19}$} V. P. Gusynin,
		Nucl. Phys. {\bf B 333}, 296 (1990);
\item{} P. A. Carinhas and S. A. Fulling,
		in: {\it Asymptotic and computational  analysis}, Proc. Conf.
		in
		 Honor of Frank W. J. Olver's 65th birthday, ed.  R.  Wong
		(Marcel Dekker, New York, 1990) p. 601.
\item{$^{20}$} I. G. Avramidi,
		{\it Covariant methods for calculating the low-energy effective
		action in quantum field theory and quantum gravity},
		University of Greifswald (1994), gr-qc/9403036;
\item{} I. G. Avramidi,
		{\it  New algebraic methods for calculating the heat kernel
		and the effective action in quantum gravity and gauge
		theories}, gr-qc/9408028, in: {\it `Heat Kernel Techniques
		and Quantum Gravity'\ (Winnipeg Conference)}, Discourses in
		Mathematics and Its Applications,  No.~4, ed.\ by  S.~A.
		Fulling, Texas A\&M University, (College Station, Texas, 1995),
		to appear;
\item{} I. G. Avramidi,
		Phys. Lett. {\bf B336}, 171 (1994).
\item{$^{21}$} I. G. Avramidi,
		Phys. Lett. {\bf B 305}, 27 (1993).

\item{$^{22}$} J. S. Dowker,
		Ann. Phys. (USA) {\bf 62}, 361 (1971);
		J. Phys. {\bf A 3}, 451 (1970);
\item{} A. Anderson and R. Camporesi,
		Commun. Math. Phys. {\bf 130}, 61 (1990);
\item{} R.  Camporesi,
		Phys. Rep. {\bf 196}, 1 (1990);
\item{} N. E. Hurt,
		{\it Geometric quantization in action:  applications of
		harmonic
		analysis in quantum statistical mechanics and quantum field
		theory}, (D. Reidel Publishing Company,  Dordrecht, Holland,
		1983).

\item{$^{23}$} E. S. Fradkin and A. A. Tseytlin,
		Nucl. Phys. {\bf B 234}, 472 (1984);
\item{} I. L. Buchbinder, S. D. Odintsov, I. L. Shapiro,
		{\it Effective action in quantum gravity}, (IOP Publishing,
		Bristol, 1992);
\item{} G. Cognola, K. Kirsten and S. Zerbini,
		{\it One-loop effective potential on hyperbolic manifolds},
		Trento University (1993);
\item{}  I. G. Avramidi,
		J. Math. Phys. {\bf 36}, 1557 (1995).
\item{$^{24}$} J. A. Zuk,
		Phys. Rev. {\bf D 33}, 3645 (1986);
\item{}     V. P. Gusynin and V. A. Kushnir,
		Class. Quantum Grav. {\bf 8}, 279 (1991).

\item{$^{25}$} H. S. Ruse, A. G. Walker, T. J. Willmore,
		{\it Harmonic spaces}, (Edizioni Cremonese, Roma (1961));
\item{} J. A. Wolf,
		{\it Spaces of constant curvature,}
		(University of California, Berkeley, CA, 1972);
\item{}  B. F. Dubrovin, A. T. Fomenko and S. P. Novikov,
		{\it The Modern  Geometry: Methods and  Applications},
		(Springer, N.Y. 1992).

\item{$^{26}$} M. Takeuchi, {\it Lie Groups II, in Translations of Mathematical
Monographs}, vol. 85, (AMS, Providence, 1991), p.167

\item{$^{27}$} H. D. Fegan,
		J. Diff. Geom. {\bf 18}, 659 (1983).
\item{$^{28}$} M. R. Brown and M. J. Duff,
		Phys. Rev. {\bf D11}, 2124 (1975).
\item{$^{29}$} B. L. Hu and D. J. O'Connor,
		Phys. Rev. {\bf D 30}, 742 (1984).

\bye